\documentclass[12pt]{svmult}
\usepackage[top=1in, bottom=1.25in, left=1.25in, right=1.25in]{geometry}

\usepackage{graphicx}        
\usepackage{multicol}        
\usepackage[bottom]{footmisc}

\usepackage{enumitem}
\usepackage{amssymb}
\usepackage{amsmath}
\usepackage{setspace}
\usepackage{framed}
\usepackage{xcolor}
\usepackage[numbers,sort]{natbib}

\def\rem#1 {{\em#1}}
\def\ddt#1 {\frac{\mathrm{d}#1}{\mathrm d t}}

\makeindex             

\begin{document}

\title*{Effects of additive noise on the stability of glacial cycles}
\author{Takahito Mitsui and Michel Crucifix}
\institute{Takahito Mitsui \at Universit\'e catholique de Louvain, Earth and Life Institute, George Lema\^\i tre Centre for Earth and Climate Research, BE-1348 Louvain-la-Neuve, Belgium\\ 
\email{takahito321@gmail.com}
\and Michel Crucifix \at Universit\'e catholique de Louvain, Earth and Life Institute, George Lema\^\i tre Centre for Earth and Climate Research, BE-1348 Louvain-la-Neuve, Belgium,\\
Belgian National Fund of Scientific Research, Rue d'Egmont, 5 BE-1000 Brussels, Belgium\\
\email{michel.crucifix@uclouvain.be}
}
%
%
\maketitle

\abstract*{It is well acknowledged that the sequence of glacial-interglacial cycles is paced by the astronomical forcing. However, how much is the sequence robust against natural fluctuations associated, for example, with the chaotic motions of atmosphere and oceans? In this article, the stability of the glacial-interglacial cycles is investigated on the basis of simple conceptual models. Specifically, we study the influence of additive white Gaussian noise on the sequence of the glacial cycles generated by stochastic versions of several low-order dynamical system models proposed in the literature. In the original deterministic case, the models exhibit different types of attractors: a quasiperiodic attractor, a piecewise continuous attractor, strange nonchaotic attractors, and a chaotic attractor. We show that the combination of the quasiperiodic astronomical forcing and additive fluctuations induce a form of temporarily quantised instability. More precisely, climate trajectories corresponding to different noise realizations generally cluster around a small number of stable or transiently stable trajectories present in the deterministic system. Furthermore, these stochastic trajectories may show sensitive dependence on very small amounts of perturbations at key times. Consistently with the complexity of each attractor, the number of trajectories leaking from the clusters may range from almost zero (the model with a quasiperiodic attractor) to a significant fraction of the total (the model with a chaotic attractor), the models with strange nonchaotic attractors being intermediate. Finally, we discuss the implications of this investigation for research programmes based on numerical simulators.  }

\abstract{It is well acknowledged that the sequence of glacial-interglacial cycles is paced by the astronomical forcing. However, how much is the sequence robust against natural fluctuations associated, for example, with the chaotic motions of atmosphere and oceans? In this article, the stability of the glacial-interglacial cycles is investigated on the basis of simple conceptual models. Specifically, we study the influence of additive white Gaussian noise on the sequence of the glacial cycles generated by stochastic versions of several low-order dynamical system models proposed in the literature. In the original deterministic case, the models exhibit different types of attractors: a quasiperiodic attractor, a piecewise continuous attractor, strange nonchaotic attractors, and a chaotic attractor. We show that the combination of the quasiperiodic astronomical forcing and additive fluctuations induce a form of temporarily quantised instability. More precisely, climate trajectories corresponding to different noise realizations generally cluster around a small number of stable or transiently stable trajectories present in the deterministic system. Furthermore, these stochastic trajectories may show sensitive dependence on very small amounts of perturbations at key times. Consistently with the complexity of each attractor, the number of trajectories leaking from the clusters may range from almost zero (the model with a quasiperiodic attractor) to a significant fraction of the total (the model with a chaotic attractor), the models with strange nonchaotic attractors being intermediate. Finally, we discuss the implications of this investigation for research programmes based on numerical simulators.  }

\section{Introduction \label{sect:intro}}
Analyses of marine sediments and ice core records show, among others, that glacial and interglacial periods alternated over the last three million years {\cite{shackleton76, lisiecki05lr04}.
These are major climate changes. 
The last glacial maximum that occurred about 21,000~years ago was characterised by extensive ice sheets over large fractions of North America, the British Isles and Fennoscandia. Sea-level was about 120~m below the present-day, and the CO$_2$ concentration was about 90~ppm lower than its typical pre-industrial value of 280~ppmv \cite{lambeck01sealevel, petit99}. 

 \begin{figure}[ht]
\begin{center}
  \includegraphics[scale=0.45,angle=270]{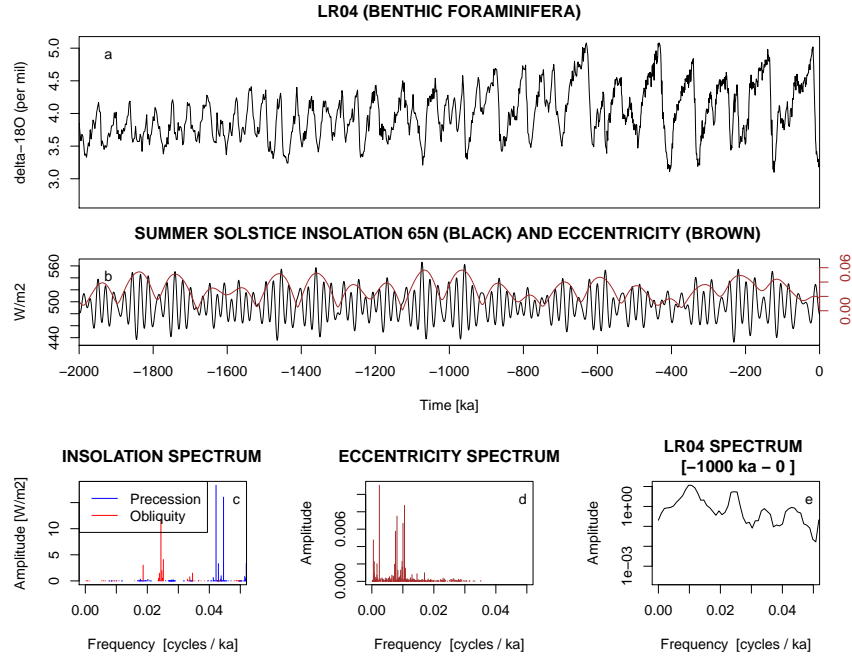}
\caption{
(a) Reconstructed climate variations over the last 2 million years (1~ka stands for 1,000 years) inferred from deep-sea  organisms, specifically benthic foraminifera \cite{lisiecki05lr04}, along with (b) the variations in incoming solar radiation at the summer solstice at 65$^\circ$ N (black), a classical measure of astronomical forcing computed here following the BER90 algorithm \cite{Berger91aa}. The spectrum of insolation is (c), with components arising from climatic precession and obliquity computed following \cite{berger78,bergerl90}. Eccentricity (figure (b), brown) is the modulating envelope of precession, and its spectrum is given in (d) \cite{berger78,bergerl90}. (e)  multi-taper estimate of the LR04 spectrum (last 1 million years only) estimated using multi-taper method \cite{Thompson82aa} obtained using the SSA-MTM toolkit \cite{Vautard92aa} with default parameters. 
} \label{fig:data-insol}
\end{center}
\end{figure}

The so-called ``LR04" time series \cite{lisiecki05lr04} is a compilation of records of Oxygen isotopic ratio $\delta ^{18}$O recorded in deep-sea organisms. This is representative of the succession of glacial-interglacial cycles (Fig.~\ref{fig:data-insol}).   
It may be seen that the temporal signature of glacial-interglacial cycles has evolved through time: their amplitude increased gradually, and about 1 million years ago, their period settled to about 100 ka\footnote{In the following, 1~ka = 1,000 years and 1~Ma = 1,000~ka.}.  The four latest cycles are particularly distinctive, with a gradual glaciation phase extending over about 80~ka, and a deglaciation over 10-20 ka \cite{broecker70, petit99}.

    Glacial cycles emerge from a complex interplay of various physical, biogeochemical and geological processes, and it is hoped that their detailed analysis will yield information on the stability of the different components of the climate system. 
    It has also become clear that glacial cycles are submitted to an external control. 
    In particular, the timing of deglaciations is statistically related to quasiperiodic changes in Earth's orbit and obliquity \cite{Raymo97aa,Huybers05obliquity,  Lisiecki10aa}. 
    One of the key mechanisms of this control is that changes in Earth's orbit and obliquity influence
    the seasonal and spatial distributions of the incoming solar radiation (insolation) at the top of the atmosphere. 
    Specifically, the insolation of at a given time of the year at a given latitude is approximately a linear function of $e\sin\varpi$, $e\cos\varpi$, and $\varepsilon$, where $e$ is the Earth orbit eccentricity, $\varpi$ the longitude of the perihelion, and $\varepsilon$ the Earth's obliquity. The quantity $e\sin\varpi$ is sometimes called the climatic precession parameter \cite{Berger94aa}. In turn, 
    summer insolation at high latitude controls the mass balance of snow over the years, and affects thus the growth of ice sheets \cite{milankovitch41, weertman76, berger88, Abe-Ouchi13aa}. This is, however, probably not the only mechanism of astronomical control on ice ages \cite{Ruddiman06ab}. 
   
   A long-standing puzzling fact is that the Fourier spectrum of insolation changes mainly contains power around 20 and 40~ka \cite{berger78}, while the  spectrum of the slow fluctuations of climate shows a concentration of power around 100~ka (Fig.~\ref{fig:data-insol}, see also \cite{hays76,  Wunsch03spectral,BOLTON95aa,Rial13aa}).  The periods of $19-23$~ka arise from precession (the rotation period of $\varpi$) while $40$~ka is the dominant period of obliquity. 
    In fact, periodicities around 100 ka do appear in astronomical forcing, but somewhat indirectly \cite{Berger05aa}. 
    In particular, eccentricity, which  modulates the amplitude of climatic precession, is characterised by a spectrum with periods around 100 and 413~ka \cite{berger77,Berger91aa}.
    The correspondence between the 100 ka period of eccentricity and the duration of ice ages was noted early on \cite{hays76, imbrie93}, and statistical analysis of the timing of ice ages indicates that this correspondence is probably not fortuitous  \cite{imbrie93,Raymo97aa,Lisiecki10aa}.

These observations lead us to  an interesting problem: is it possible to predict the effects of the astronomical forcing on the ice ages without full knowledge of the detailed physical mechanisms? 
Specifically, is it possible to determine whether the sequence of ice ages is tightly controlled by the astronomical forcing or whether, to the contrary, this sequence is highly sensitive to small fluctuations? 

The strategy proposed here relies on the analysis of low-order dynamical systems.
Over the years, numerous models have been proposed to explain, on the one hand, the relationship between ice ages and astronomical forcing, and, on the other hand, their specific saw-tooth temporal structure. 
A full review is beyond the scope of the present study, and we quote here some potential dynamical mechanisms, which may be relevant for modelling of ice ages:

\begin{enumerate}[label=(\alph*),itemsep=5pt]
  \item 
    Considerations on the geometry of ice sheets \cite{weertman76} suggest that a positive insolation anomaly may be proportionally more effective than a negative one. In terms of system dynamics, one may say that the forcing is transformed nonlinearly by the climate system (see also Fig.~8 of \cite{Ruddiman06ab} for other nonlinear effects). Recall that precession exerts a significant control on the seasonal distribution of insolation, and that the modulating envelope of the precession signal is eccentricity. A nonlinear transformation of the astronomical forcing is thus a simple mechanism by which the spectrum of eccentricity can make its way towards the spectrum of climate variations.
The spectrum of eccentricity includes the sought-after 100-ka period, but it is also dominated by a 413-ka period, and the latter is not observed in the benthic record. This  was sometimes referred to as the 400-ka enigma \cite{Ganopolski12aa, Rial13aa}. 
  \item 
    Results from numerical modelling suggest that the ice-sheet-atmosphere system  may present several stable states for a range of insolation forcings \cite{weertman76,  Calov2005Transient-simul,  Crucifix11aa, Abe-Ouchi13aa}. 
    Transitions between these states may be triggered deterministically (by the forcing), stochastically, or by a combination of both. 
    In early studies \cite{BENZI82aa, NICOLIS82ab}, it was suggested that 100-ka ice ages may emerge through a mechanism of stochastic resonance, in which noisy fluctuations may amplify the small eccentricity signal. This proposal is relevant to the present context because it is the first one conferring an explicit role to fluctuations, but the the stochastic resonance theory of ice ages is incomplete because it does not consider explicitly the direct effects of precession and obliquity, nor does it explain the saw-tooth shape of glacial cycles.
  \item 
    The 100~ka cycles may arise  as self-sustained (or excitable) oscillations, which itself emerge from nonlinear interactions between different Earth system components.
    These components can be ice sheets (and underlying lithosphere), deep-ocean and carbon cycle dynamics.
    Following this approach, the effect of astronomical forcing on climate may be understood in terms of the general concept of synchronisation  \cite{ashkenazy06phase, tziperman06pacing, De-Saedeleer13aa}, of which the forced van der Pol oscillator may constitute a paradigm. 
    Saltzman et al. \cite{saltzman88, saltzman90sm} published a number of low-order dynamical systems consistent with this interpretation, but  see also \cite{paillard04eps,  ashkenazy06phase, Gildor01ab} for alternatives. 
%
\item 
  If a process of nonlinear resonance occurs, 100-ka glacial cycles  may also be  obtained even if the corresponding autonomous system does not have any internal period near 100~ka. 
  A classical example of nonlinear resonance is the Duffing oscillator \cite{Kanamaru:2008}, which was recently suggested as a possible basis for the investigation of ice age dynamics \cite{Daruka15aa}. 
  While, in principle, nonlinear resonance with additive forcing may be sufficient to generate combination of tones (see also \cite{letreut83}), the expression of a dominant 100-ka cycle in response to the astronomical forcing is best obtained with multiplicative forcing (\cite{letreut83, Huybers09aa, Daruka15aa}). 
  \item 
  Ice age cycles may also be obtained in a more ad-hoc way: for example  by resorting to  a discrete variable that changes states following threshold rules involving the astronomical forcing \cite{paillard98}, or by postulating an adequate bifurcation structure in the climate-forcing space \cite{Ditlevsen09aa}.
\end{enumerate}

Naturally, different models exhibit different dynamical properties.
On this subject, it was observed that models explaining ice ages as the result of self-sustained oscillations subjected to the astronomical forcing generally display the properties of strange nonchaotic attractors \cite{Mitsui14aa, Crucifix13aj, Mitsui15ad}. 

Strange nonchaotic attractors (SNAs) may appear when nonlinear dynamical systems are forced by quasiperiodic signals \cite{Grebogi84aa,Kaneko84aa}, such as the astronomical forcing.
Unlike chaotic systems, the trajectory of such systems is typically robust against small fluctuations in the initial conditions. 
In particular, their largest Lyapunov exponent $\lambda$ is nonpositive. 
However, the attractor itself, or its stroboscopic section at one of the periodic components of the forcing, is a geometrically strange set. Comprehensive reviews on SNAs are available in \cite{Prasad01aa, Feudel06aa}. 
The strange geometry of SNAs is related to the existence of repellers of measure zero embedded in the attractor \cite{Sturman00aa}. Thus, there will be times at which the orbits on SNAs are arbitrarily close to the orbits on the repellers. As a result, it is shown that the trajectories generated by models with SNAs may have sensitive dependence on parameters \cite{Nishikawa96aa}.
They may also show sensitive dependence on dynamical noise \cite{Khovanov00aa}. 

The concept of SNA was introduced in ice age theory on the basis of low-order deterministic models \cite{Mitsui14aa, Crucifix13aj, Mitsui15ad}, but such models are naturally gross simplifications of the complex climate system. In this context, one  step forward is to enrich the dynamics with stochastic parameterisations, in order to represent the effects of fast climate and meteorological processes \cite{Hasselmann76,  saltzman90sm, Penland03aa}.
As already mentioned, stochastic parameterisations were introduced in the palaeoclimate context as an element of the stochastic resonance theory  \cite{BENZI82aa, NICOLIS82ab}. Stochastic processes were also considered in simple ice age models to illustrate the process of synchronisation \cite{Nicolis87aa, tziperman06pacing, Crucifix11aa}, to induce stochastic jumps between different stable equilibria \cite{Ditlevsen09aa}, or to induce coherence resonant oscillations \cite{Pelletier03aa}

Our specific objective is here to study the effect of dynamical noise (or system noise) on the robustness of palaeoclimate trajectories generated by simple stochastic models forced by the astronomical forcing, and explain differences among the models by reference to the properties of the attractors displayed by the deterministic counterparts of these models. To this end, we consider four simple models known to exhibit different kinds of attractors, and consider the effects of additive white Gaussian noise on the simulated sequence of ice ages. 

Modelling fast meteorological and climatic processes by additive Gaussian noise may be oversimplified though it is frequently used in studies of ice ages \cite{BENZI82aa, NICOLIS82ab, saltzman90sm, tziperman06pacing, Ditlevsen09aa}. In the studies of millennial-scale climate changes (so-called Dansgaard Oeschger events), Ditlevsen (1999) employs the $\alpha$-stable noise, which is characterized by a fat-tailed density distribution \cite{Ditlevsen99aa}, and mathematical methods to identify the $\alpha$-stable noise in time series have been developed (see Gairing et al. in this volume). Colored multiplicative noises were also introduced in box models of thermohaline circulation \cite{Timmermann00aa}. Such non-Gaussian or colored multiplicative noises can be relevant to ice age dynamics. However, here we focus on the effects of additive white Gaussian noises as a first step to examine the stability of ice age models.

\section{Methods}
The astronomical forcing $F(t)$ is represented here as a linear combination of forcing functions associated with obliquity and climatic precession.
For consistency with previous works we use the summer-solstice standardised (zero-mean) insolation approximated as a sum of 35 periodic functions of time $t$, as in \cite{De-Saedeleer13aa} and \cite{Mitsui14aa} (see also \cite{Crucifix11aa}),
\begin{equation}
F(t)=\frac{1}{A} \sum _{i=1}^{35} [s_i \sin \omega _i t+c_i \cos \omega _i t],
\end{equation}
where the scale factor $A$ is set to 11.77~W/m$^2$ for the CSW model mentioned below and 23.58~W/m$^2$ for the other models.

The current study is focused around four previously published conceptual models whose parameter values are listed in Appendix:
\begin{itemize} 
  \item The model introduced by Imbrie and Imbrie \cite{imbrie80} (I80) is one dimensional ordinary differential equation in which the climate-state $x$ (a measure of the global ice volume loss) responds to the astronomical forcing $F(t)$ as follows:
    \begin{equation} \label{eq:i98}
      \tau \frac{\mathrm {d} x}{\mathrm{d}t} = \left\{
\begin{split}
&(1+b)(F(t)-x)\,\,\,\text{if}\,\,\,F(t)\geq x\\
&(1-b)(F(t)-x)\,\,\,\text{if}\,\,\,F(t)<x.
\end{split}
\right. 
    \end{equation}
    The additional condition $x \ge 0$ expressed in \cite{imbrie80} is omitted in this study for simplicity. 
  \item The P98 model \cite{paillard98} is a hybrid dynamical system defined as follows:
    \begin{equation}\label{eq:p98}
      \frac{\mathrm {d} x}{\mathrm{d}t} = \frac{x_R-x}{\tau_R} - \frac{\tilde F(t)}{\tau_F},
    \end{equation}
    where $x$ is the global ice volume, and the relaxation time $\tau_R$ and the relaxed state $x_R$ vary discretely between $R=i$, $R=g$, and $R=G$ according to the following transition rules: 
    the transition $i\rightarrow g$  is triggered when $F(t)$ falls below a threshold $i_0$; the transition  $g\rightarrow G$  is triggered with $x$ exceeds $x_\mathrm{max}$, and $G\rightarrow i$ when 
    $F(t)$ exceeds a threshold $i_1$. 
    Note also that the forcing function used in Eq.~(\ref{eq:p98}) is a truncated version of actual insolation (a nonlinear effect), computed as follows:
    \begin{equation}
      \tilde F(t) = \frac{1}{2}\left( F(t) + \sqrt{4 a^2 + F^2(t)} \right) . \label{eq:truncation}
    \end{equation}
  \item The SM90 model \cite{saltzman90sm} is a representation of nonlinear interactions between three components of the Earth system: continental ice volume ($x$), CO$_2$ concentration ($y$) and deep-ocean temperature ($z$): 
\begin{eqnarray*}
\tau\ddt{x} &=& -x - y - v\, z - u F(t),\\
\tau\ddt{y} &=& -p \,z + r \, y + s \, z^2  -w\,yz- z^2 y, \\
\tau\ddt{z} &=& -q (x + z).
\end{eqnarray*}
The forcing is additive, and the nonlinearity introduced in the second component of the equation induces limit cycle dynamics when $u=0$. 
  \item The HA02 model: This is the same as SM90, but Hargreaves and Annan  (2002) \cite{hargreaves02} estimated the parameter values of SM90 using a data assimilation technique.
  \item The CSW model \cite{Crucifix11aa, Crucifix12aa, De-Saedeleer13aa} is in fact a forced van der Pol oscillator, used as a simple example of slow-fast oscillator with parameters calibrated such as to reproduce the ice ages record:
\begin{eqnarray*}
 \tau  \frac{\mathrm{d} x}{\mathrm{d}\,t} &=&  -  \left( \gamma F\left( t \right) + \beta + y \right), \\
 \tau  \frac{\mathrm{d} y}{\mathrm{d}\,t} &=& \alpha( y - y^3/3 + x),
\end{eqnarray*}
where $x$ is the global ice volume, and $y$ is a conceptual variable introduced to obtain a self-sustained oscillation in the absence of forcing ($\gamma =0$).
\end{itemize}

Referring to the model categories outlined in section \ref{sect:intro}, I80 belongs to category (a), SM90, HA04, and CSW to category (c), and P98 to category (e). P98 also has the particularity of being a hybrid dynamical system involving discontinuous thresholds and discrete variables, unlike the other models studied here. 

For each model, denote $\mathbf{x}$ the vector of all the climate state variables. The system equations are
\begin{equation}
\ddt{\mathbf{x}} =\mathbf{f}(\mathbf{x},\,F(t)).\label{eq:formal}
\end{equation}
If we introduce phase variables $\theta _i (t) = \omega _i t\,\,(\mbox{mod}\,2\pi)$ ($i=1, 2, ..., 35$), Eq.~(\ref{eq:formal}) can be written in a skew-product from:
\begin{eqnarray}
\ddt{\mathbf{x}}
&=&\tilde{\mathbf{f}}(\mathbf{x},\,\theta ),\\
\ddt{\theta } &=& \omega,
\end{eqnarray}
where $\mathbf{\theta }=(\theta _1,\,\theta _2,\,...,\,\theta _{35})$ and $\mathbf{\omega }=(\omega _1,\,\omega _2,\,...,\,\omega _{35})$.
We consider the attractor of each model in the extended phase space ($\mathbf{\theta },\,\mathbf{x}$). As time $t$ elapses enough from the initial time $t_0$, trajectories approach the attractor (cf. \cite{Grebogi84aa} for a definition of attractor in this particular context). 
To see a qualitative difference between the categories, we show the attractors of each model for a simplified forcing $F_s(t)= \frac{1}{A_{ \{1,3,4\} }} \sum _{i=1, 3, 4}(s_i\sin \omega _it +c_i \cos \omega _i t)$, where indices $i=1,\,3,\,4$ and parameter $A_{\{1,3,4\}}$ are consistent with \cite{Mitsui14aa}. 
The attractors of each model for the simplified forcing $F_s(t)$ are shown in Fig.~\ref{fig:attractors}. To visualise the high-dimensional attractors, the state points are plotted in a three-dimensional space of ($\theta _3/2\pi , \theta _4/2\pi , x$) at a regular time interval of $12 \pi /\omega _1$ (the so-called stroboscopic section). For each model, these plots show the relationship between the phases of the astronomical forcing and the variable representing ice volume.
 \begin{figure}[ht]
\begin{center}
\includegraphics[scale=0.8,angle=0]{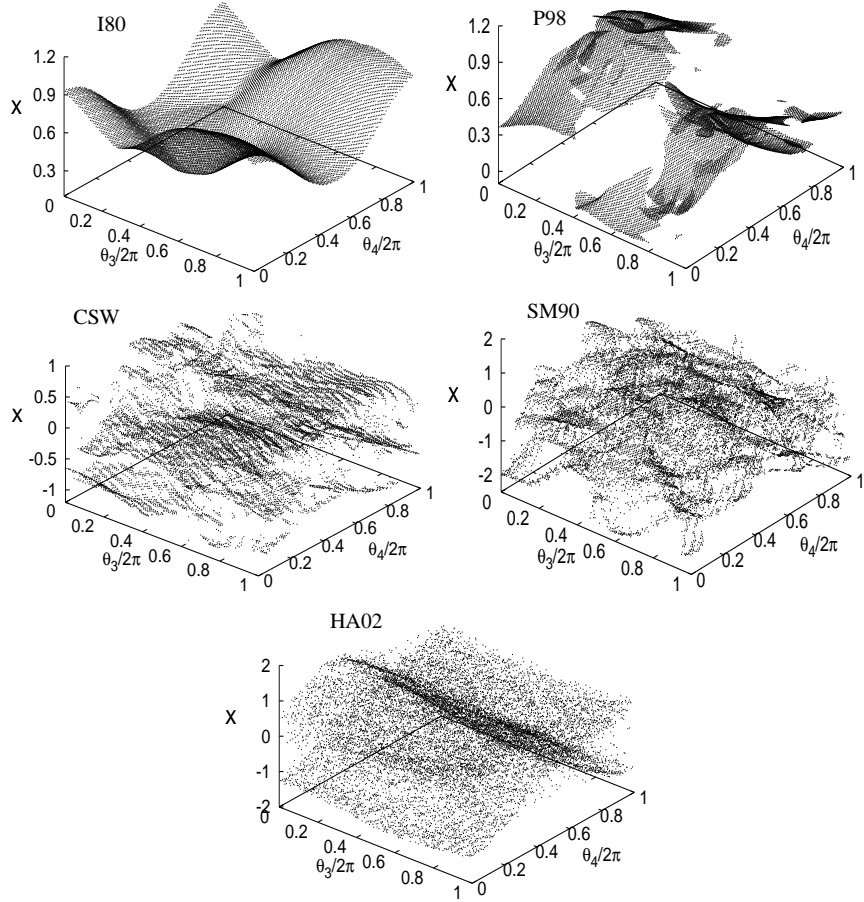}
\caption{
Attractors of each model for a simplified astronomical forcing $F_s(t)$. To visualise the high-dimensional attractors, the state points are plotted in a three-dimensional space of ($\theta _3/2\pi , \theta _4/2\pi , x$) at a regular time interval of $12 \pi /\omega _1$. Transients are removed.
} \label{fig:attractors}
\end{center}
\end{figure}

The different geometries associated with these models can be readily identified (Fig.~\ref{fig:attractors}) (see also \cite{Mitsui14aa}). 
The I80 model has a smooth attractor. More specifically, the stroboscopic section is a smooth surface, and the time evolution of a trajectory is quasiperiodic. 
The stroboscopic section of P98 appears as piecewise smooth, which is not surprising considering the fact that the equations are linear, expect for the state transitions at threshold values of state or insolation. 
Consistently with earlier analysis \cite{Mitsui14aa}, the CSW and SM90 exhibit SNAs. The stroboscopic sections appear discontinuous almost everywhere. It can qualitatively be discerned that the sections of CSW and SW90 are more organised than the section of HA02, which is known to be chaotic. Recall again that SM90 and HA02 are the same equations, but with different parameters, and the two regimes are separated by a transition from SNA (negative largest Lyapunov exponent $\lambda$) to chaos (positive $\lambda$). 

Stochastic versions are now defined for each model. 
As ice volume $x(t)$ is the only state variable common among all the models, dynamical noise is added only to the equation of $x(t)$ to allow us to compare models.  The equation for the ice volume $x(t)$ is thus schematically written as:
\begin{displaymath}
  \mathrm{d} x = f(\mathbf{x},t) \mathrm{d}t +D\,\mathrm{d}W(t),
\end{displaymath}
where $W(t)$ is the Wiener process, and $D$ is the noise intensity, and $f(\mathbf{x},t)$ denotes the derivative $\frac{\mathrm{d} x}{\mathrm{d}t}$ entering the corresponding deterministic model.
To account for the fact that the typical size of ice volume variations is different among models, 
we introduce the scaled noise intensity $\sigma =D/ L$, where $L$ is the standard deviation of ice volume $x(t)$ for each model calculated during [-700 ka, 10 Ma] in the absence of noise (see Appendix for the value of $L$ in each model).
The initial time is set at $t_0=-20$~Ma for SM90 and HA02, and $t_0=-2$~Ma for the other models to discard initial transients \cite{Mitsui14aa}. 
Each model is integrated from $t=t_0$ to $t_s=-700$~ka without dynamical noise and then integrated with dynamical noise $D\,\mathrm{d}W (t)$ from $t=t_s$ to $t=t_e$.  We simulate $N(=200)$ trajectories corresponding to different realizations of the Wiener process $W (t)$. All models are integrated using the stochastic Heun method with a time step of $0.001$~ka \cite{Greiner88aa}. 

The ensemble of ice volumes $\{x_i(t): i=1, ..., N\}$ disperses due to different noise realizations. 
Twenty sample trajectories generated by CSW model for $\sigma =0.002$ are shown in Fig.~\ref{fig:demo}.  
As earlier noted \cite{De-Saedeleer13aa, Crucifix13aj,  Mitsui14aa}, the astronomical forcing induces a form of synchronisation, such that the different noisy trajectories tend to remain clustered. 
There are however times at which clusters break apart, yielding a temporarily more disorganised picture. 
This is the behaviour that we wish to characterise more systematically.

\begin{figure}[ht]
\begin{center}
\includegraphics[scale=0.4,angle=270]{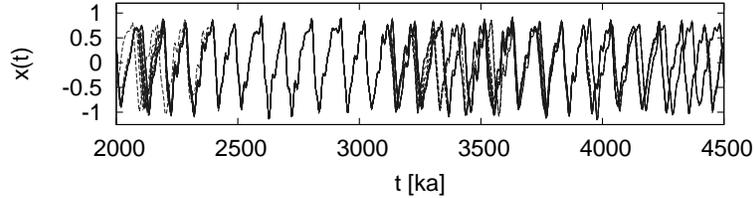}
\caption{
Effect of dynamical noise in the CSW model. Twenty sample trajectories of ice volume $x(t)$ corresponding to different noise realizations with $\sigma =0.002$.  
} \label{fig:demo}
\end{center}
\end{figure}

To this end, the dispersions of trajectories in the models are analyzed by using the following three quantities:

\begin{itemize}
\item {\it Size of dispersion of ice volume $x_i(t)$ at a time instant $t$, $S(t)$, compared to the typical size of ice volume variation $L$} is given by
\begin{equation}
S(t)=\left( \frac{1}{N} \sum _{i=1}^N [x_i(t)-\langle x_i(t)\rangle ]^2 \right)^{1/2}/L,
\end{equation}
where $\langle x_i (t)\rangle =\frac{1}{N}\sum _{i=1}^N x_i(t)$ is the ensemble average of $N=200$ trajectories. The size of dispersion $S(t)$ may be used as a measure of dynamical complexity induced by noise \cite{Khovanov00aa}.

\item {\it Number of large clusters at a time instant $t$, $N_{LC}(t)$.}
For every time $t$, the $N$ model states associated with the $N$ sample solutions of the stochastic differential equation are clustered. The metric used for clustering the states of I80, CSW, SM90, and HA02 is simply the Euclidean distance in, respectively, the 1-, 2-, and 3-D spaces of climate variables $\mathbf{x}$. Different approaches may be imagined for P98. One pragmatic and sufficiently robust solution is to define an auxiliary variable $y$ taking values $1$, $2$, or $3$ for states $i$, $g$, and $G$, respectively, and define the Euclidean distance in the $(x,y)$ space.
With this distance at hand, clusters are defined using the following iterative algorithm, similar to \cite{De-Saedeleer13aa}.
\begin{enumerate}
\item 
  Define $\mathbb{I}  = \{1, 2, \ldots , N\}$ 
  the indices of the ensemble to be clustered, and call $\mathbf{x}_{i} (t)$ the $i^\mathrm{th}$ model state.
\item Set $j=1$ and repeat the following steps until $\mathbb{I}$ becomes empty:
  \begin{enumerate}
  \item Call $\mathbf{x}^\star (t)$ the model state corresponding to one of the members of ensemble $\mathbb{I}$  
  \item Define $\mathbb{I}_j$, the ensemble of indices $\{i \in \mathbb{I}\,:\,||\mathbf{x}^\star (t) - \mathbf{x}_i (t)||<\epsilon \}$ 
  \item Update $\mathbb{I} = \mathbb{I} \setminus \mathbb{I}_j$
  \item Increment $j = j+1$ if $\mathbb{I}\neq \emptyset$
  \end{enumerate}
\item The clusters are the $\{\mathbb{I}_j\}$.
\end{enumerate}
We use $\epsilon=0.7$ for SM90 and HA02 and $\epsilon=  0.4$ for the other models. 
The number of \textit{large} clusters, $N_{LC}(t)$, is defined as the number of clusters with at least ten members. 

\item {\it Finite-time Lyapunov exponent $\lambda _T(\mathbf{x}(t),\delta \mathbf{x}(t_0))$ for a time interval [$t,t+T$]} is defined as in \cite{Eckhardt93aa}:
\begin{equation}\label{eq:lyap}
\lambda _{T}(\mathbf{x}(t),\delta \mathbf{x}(t_0))=\frac{1}{T} \ln \frac{|\delta \mathbf{x}(t+T)|}{|\delta \mathbf{x}(t)|},
\end{equation}
where $\delta \mathbf{x}(t)$ is a vector representing an infinitesimal deviation from a reference trajectory of the climate state, $\mathbf{x}(t)$. The vector $\delta \mathbf{x}(t)$ is given as a solution of the linearlized equation of the original dynamical system. The finite-time Lyapunov exponent $\lambda _T(\mathbf{x}(t),\delta \mathbf{x}(t_0))$ gives the rate of exponential divergence of nearby orbits from the reference trajectory $\mathbf{x}(t)$ during the time interval [$t,\,t+T$]. Typical initial deviations $\delta \mathbf{x}(t_0)$ give a same value for each trajectory $\mathbf{x}(t)$ for $t\gg t_0$. For simplicity, we denote $\lambda _T(\mathbf{x}(t),\delta \mathbf{x}(t_0))$ by $\lambda _T(t)$, but note that $\lambda _T(t)$ still depends on $\mathbf{x}(t)$. A positive (negative) value of $\lambda _T(t)$ indicates temporal instability (stability) of a trajectory in the time interval [$t,\,t+T$]. As $T\to \infty$, it converges to the largest Lyapunov exponent $\lambda$.
\end{itemize}

\section{Results}
\subsection{Dispersions in each model}
We compare the noise sensitivity of each model by using the maximum size of dispersion $S_{\max}$ and the mean size of dispersion $S_{\mathrm{mean}}$ during the time interval [$t_s,\,t_e$]:
\begin{eqnarray*}
S_{\max} &=& \max _{t_s\leq t\leq t_e} S(t),\\
S_{mean} &=&\frac{1}{t_e-t_s} \int _{t_s}^{t_e} S(t)dt.
\end{eqnarray*}
First, we consider these quantities in a long time interval from $t_s=-700$~ka to $t_e=100$~Ma,
in order to characterise global properties of the attractor of each model. The maximum $S_{\max}$ and the mean $S_{\mathrm{mean}}$ are presented as functions of the scaled noise intensity $\sigma \in [0.001,\,0.1]$ in Figs.~\ref{fig:dispersion_summary}(a) and \ref{fig:dispersion_summary}(b).  
\begin{figure}[ht]
\begin{center}
\includegraphics[scale=0.65,angle=0]{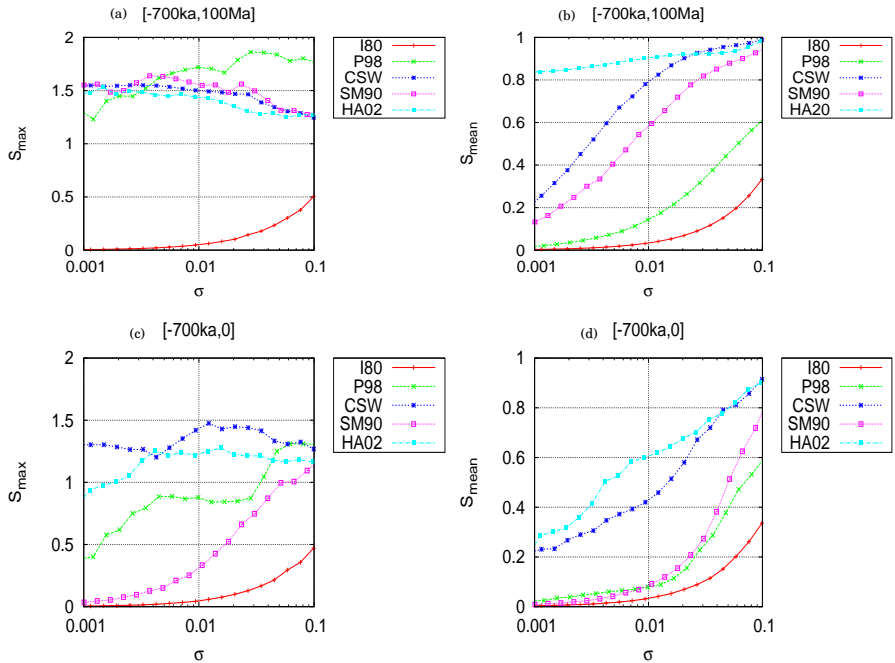}
\caption{
Comparison of noise sensitivity between the models: (a) Maximal size of dispersion $S_{\max}$ in the long time interval [-700~ka, 100~Ma]. (b) Mean size of dispersion $S_{mean}$ in the long time interval [-700~ka, 100~Ma]. (c) Maximal size of dispersion $S_{\max}$ in the time interval [-700~ka, 0]. (d) Mean size of dispersion $S_{mean}$ in the time interval [-700~ka, 0].
} \label{fig:dispersion_summary}
\end{center}
\end{figure}
The I80 model is fairly robust against dynamical noise, but the other models are highly sensitive to dynamical noise in the sense that large dispersions of trajectories, $S_{\max}\sim 1$, can be induced by extremely small noise  (e.g. $\sigma \sim 0.001$) if one waits long enough  (Fig~\ref{fig:dispersion_summary}(a)). The mean size of dispersion $S_{\mathrm{mean}}$ is relatively large in SNA models (CSW and SM90) and the chaotic model (HA02), but it is small in the P98 model (Fig~\ref{fig:dispersion_summary}(b)). These differences become less obvious for large noise $\sigma >0.1$.

We now examine the qualitative differences between the dynamics of dispersions generated by each model in particular for small dynamical noise $\sigma =0.002$.
Figures~\ref{fig:loclya_clust}(a)--\ref{fig:loclya_clust}(e) present the time series of the number of large clusters $N_{LC}(t)$ (top), the size of dispersion $S(t)$ (middle), and the finite-time Lyapunov exponent $\lambda _T(t)$ (bottom) for each model. 
\begin{figure}[ht]
\begin{center}
\includegraphics[scale=0.55,angle=0]{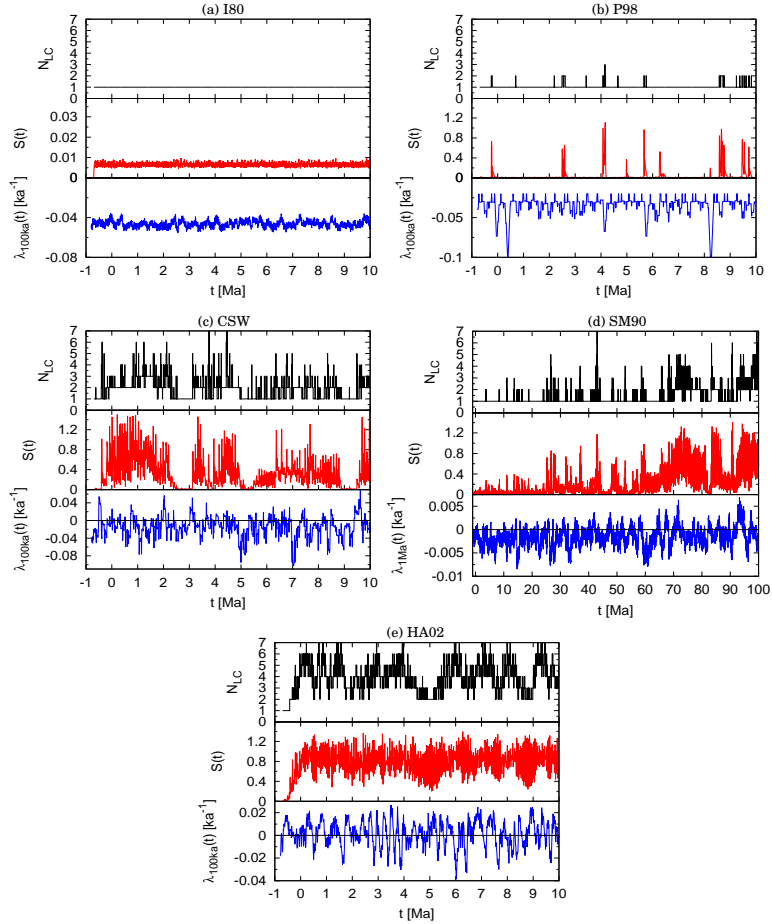}
\caption{
Responses to small dynamical noise $\sigma =0.002$ of each model: Number of large clusters $N_{LC}(t)$ (top, black). Size of dispersion $S(t)$ (middle, red). Finite-time Lyapunov exponent $\lambda _T(t)$ of the unperturbed system (bottom, blue). The averaging time $T$ for $\lambda _T(t)$ is 1~Ma for SM90 and 100~ka for the other models. Note that the time interval of panel (d) SM90 is longer than the others because of its slow evolution of dispersion.
} \label{fig:loclya_clust}
\end{center}
\end{figure}
No large dispersion appears in the I80 model. Large dispersions intermittently occur in the other nonchaotic systems (P98, CSW, and SM90) (cf. again Fig.~\ref{fig:demo}). Periods of synchronisation ($N_{LC}(t)=1$) may extend over several glacial cycles, unlike what is seen in the chaotic system (HA02) (though we note a period with two large clusters in the chaotic case, around $t=+5$~Ma). 

Let us now focus on the intermittent dispersions. In the P98 model, the episodes of large dispersion are infrequent and relatively short (typically, one or two glacial cycles) because large dispersions are caused only at thresholds and elsewhere the system is stable.
In the models with SNA (CSW and SM90), the episodes of large dispersion can last several million years in CSW and several tens of million years in SM90. These large dispersions are caused by temporal instability of the system. In fact, it may be observed that the original deterministic systems tend to have a large positive value of the finite-time Lyapunov exponent $\lambda _T(t)$ before the onsets of the dispersions (Figs.~\ref{fig:loclya_clust}(c) and \ref{fig:loclya_clust}(d) (bottom)). The finite-time Lyapunov exponent of the system under the dynamical noise behaves similarly as Figs.~\ref{fig:loclya_clust}(c) and \ref{fig:loclya_clust}(d) since the dynamical noise is small (data are not shown). 

\subsection{Order in dispersions}
In the models with SNAs (CSW and SM90), the long-lasting large dispersions are related to the existence of transient orbits with a long life time. 
The existence of transient orbits is illustrated in Fig.~\ref{fig:transient}(a) (top, blue lines) using the CSW model: stochastic trajectories are generated with $\sigma =0.002$ over the time interval $[-700\,\mbox{ka},\,0]$; then the noise is shut off and the system is integrated with the original deterministic equation. Transient orbits are excited by noise slightly before the deglaciation around $t= -400$~ka, where the finite-time Lyapunov exponent $\lambda _{200\mbox{ka}}(t)$ is temporarily positive (Fig.~\ref{fig:transient}(a), bottom). These transient orbits may then last over more than 1 million years after the cessation of dynamical noise. They are attractive, in the sense that the finite-time Lyapunov exponent $\lambda _{200\mbox{ka}}(t)$ is negative on average in time (Fig.~\ref{fig:transient}(a), bottom, blue lines). As a result, when large dispersions occur, individual trajectories may get attracted either around the trajectory corresponding to the attractor of the original deterministic system or around some pieces of transient orbits that have a long life time, as shown in Fig.~6(b).

The existence of such stable transient orbits was reported by Kapitaniak \cite{Kapitaniak92aa}, where they were termed {\it strange nonchaotic transients}. They can also be related to the notion of finite-time attractivity, and more specifically that of ($p,T$)-attractor defined by Rasmussen \cite{Rasmussen00aa} (pp.~19--20).
\begin{figure}[ht]
\begin{center}
\includegraphics[scale=0.7,angle=0]{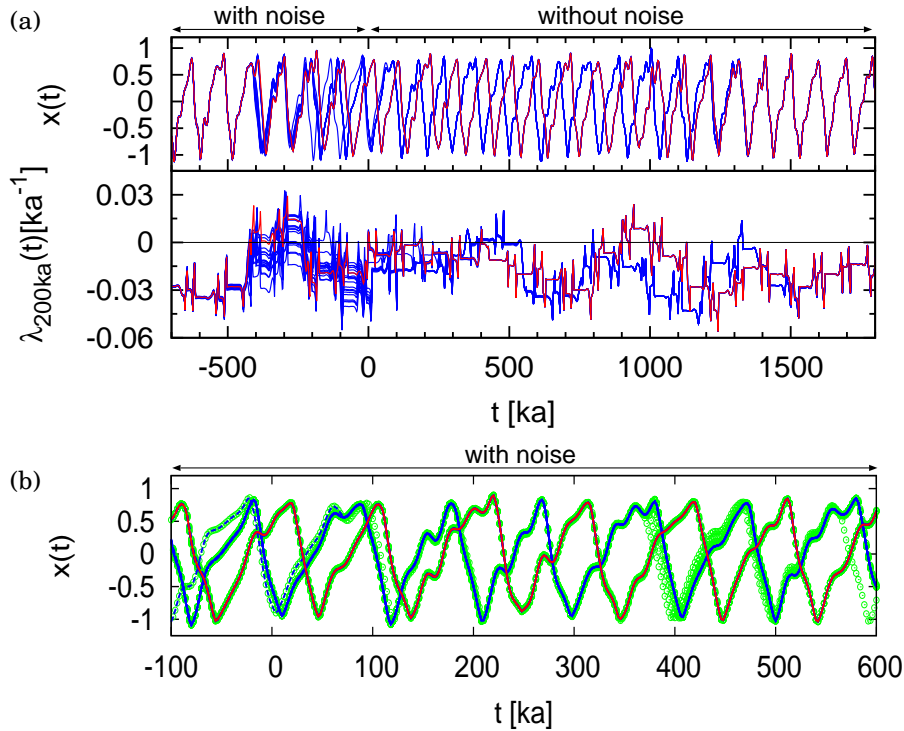}
\caption{Transient orbits with a long life time in CSW model: (a) The trajectory corresponding to the attractor of the original noiseless system (red) and some pieces of transient orbits with a long life time (blue) (top). The finite-time Lyapunov exponent $\lambda _{200\mbox{ka}}(t)$ for the trajectories in the top panel (bottom). (b) Twenty trajectories under the dynamical noise with $\sigma =0.002$ (green points). The red and blue lines are the same in the top panel in (a).  
} \label{fig:transient}
\end{center}
\end{figure}

Dispersed trajectories in the models with SNA (CSW and SM90) are more clustered than those in the chaotic model (HA02). The number of large clusters $N_{LC}(t)$ in the models with SNA (CSW and SM90) is smaller on average than that in the chaotic model (HA02), as shown in Figs~\ref{fig:loclya_clust}(c), \ref{fig:loclya_clust}(d), and \ref{fig:loclya_clust}(e). Furthermore, the number of the points outside large clusters is larger in the chaotic model, as shown in Fig.~\ref{fig:outside}. This result is intuitively reasonable: given that the largest Lyapunov exponent $\lambda$ is positive but small in the chaotic model ($\lambda \approx 0.0020$~ka$^{-1}$), we do expect trajectories to slowly travel away from the center of the clusters and thus distribute over the phase space 
more widely than in the case of strange nonchaotic models.

\begin{figure}[ht]
\begin{center}
\includegraphics[scale=0.5,angle=0]{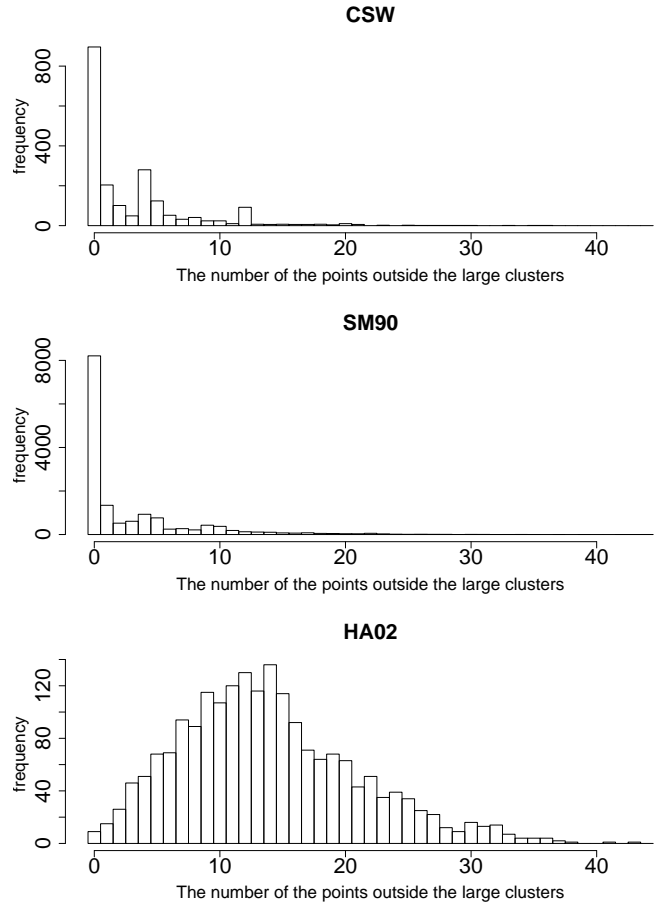}
\caption{
Frequency distribution for the number of the points outside the large clusters. These results correspond to Figs.~\ref{fig:loclya_clust}(c), \ref{fig:loclya_clust}(d), and \ref{fig:loclya_clust}(e).
} \label{fig:outside}
\end{center}
\end{figure}

\subsection{Implication in the time scale of ice ages}
The time interval focused so far, [-700~ka, 100~Ma], is quite long from the viewpoint of ice ages. Figures~\ref{fig:dispersion_summary}(c) and \ref{fig:dispersion_summary}(d) present the maximum size $S_{\max}$ and the mean size $S_{\mbox{mean}}$ of dispersions calculated in the ``short'' time interval [-700~ka, 0], where 100-ka glacial cycles took place in the history. In this short time window, the SM90 model with SNA is relatively robust against noise of $\sigma \sim O(0.01)$ because dispersions cannot evolve to the attractor size owing to its weak instability. On the other hand, the CSW model, also with SNA, is not robust against the noise of $\sigma \sim O(0.01)$ owing to its stronger temporal instability $\lambda _{\text{100~ka}}(t)\sim 0.05$ ka$^{-1}$ appeared around $t\sim -0.5$~Ma (Fig.\ref{fig:loclya_clust}(c), bottom). To assess the stability of dynamical systems models of ice ages, it is useful to know not only the type of attractor (such as quasiperiodic, piecewise smooth, strange nonchaotic, or chaotic) but also the degree of the temporal instability, which may be characterised by the local Lyapunov exponent $\lambda _{T}(t)$. 

\section{Concluding discussion}
This study shows the possibility that glacial cycles can be temporally fragile in spite of the pacing by the astronomical forcing. 

We studied the influence of dynamical noise in several models of glacial-interglacial cycles.
This analysis outlines a form of temporarily quantised instability in systems that are characterised by an SNA (CSW and SM90). 
Specifically, the systems are synchronised on the astronomical forcing, but 
large dispersions of stochastic trajectories can be induced by extremely small noise at key times when the system is temporarily unstable. 
After a dispersion event, the trajectories are organised around a small number of clusters, which may co-exist over several glacial cycles until they merge again. The phenomenon is interpreted as a noise-induced excitation of long transient orbits. Dispersion events may be more or less frequent and, depending on the amount of noise, models with SNA may have very long horizons of predictability compared to the duration of geological periods. 

Compared to this scenario, the model with a smooth attractor (I80) is always stable, i.e., large dispersion of orbits never occurs. On the other hand, the dynamics of the chaotic model (HA02) bare some similarities with the models with SNA, in the sense that trajectories cluster, i.e., at a given time the state of the system may confidently be located within a small number of regions. The difference is that there is a larger amount of leakage from the clusters, i.e., individual trajectories escape more easily from the cluster they belong to, and this reduces the predictability of such systems.  Finally, we discussed a hybrid dynamical system with a piecewise continuous attractor (P98). Owing to the discontinuity of the attractor, very small amount of noise may rarely induce significant dispersion of trajectories, and contrarily to the scenario with SNA, there are no long transients because trajectories form a single cluster rapidly. 

This analysis has some implications on the interpretation of  statistics on the relationship between the timing of ice ages and astronomical forcing. In particular, the Rayleigh statistic was used to reject a null hypothesis of independence of the phase of glacial-interglacial cycles on the components of the astronomical forcing  \cite{Huybers05obliquity, Lisiecki10aa}. Considering that  even chaotic systems show the clustering of trajectories, our results show that Rayleigh statistics, alone, are not sufficient to determine whether the sequence of ice ages is stable or not.

In this article, we did not mention the notions in random dynamical systems theory \cite{Arnold98aa,Roques13aa} so as to avoid a confusion between the classical, forward-type, definition of attractors (such as used for SNAs) and the pullback-type definition of attractors in random dynamical systems theory. However, it will be useful to formulate the present results in terms of random dynamical systems theory, where the noise-excited orbits around transient orbits may be reformulated as random fixed points. For example, in such a framework, the dynamical transition associated with a parameter change from stochastic SM90 and HA02 may be understood as a bifurcation from random fixed points to random strange attractors. Particular attention should then be paid to the nature of stochastic parameterisations and their effects on system stability. 

Taking a wider prospective, this research along with other recent works on dynamical systems of ice ages \cite{Ditlevsen09aa, Rial13aa} may provide guidance for the design and interpretation of simulations with more sophisticated models. 
One of the targets of the palaeoclimate modelling community is to simulate ice ages by resolving the dynamics of the atmosphere, ocean, sea-ice and ice sheets, coupled with adequate representations of biogeochemical processes. 
    Reasonable success has been achieved with atmosphere-ocean-ice-sheet-models forced by known CO$_2$ variation and astronomical forcing \cite{gallee92,  Ganopolski12aa,  Abe-Ouchi13aa} but simulations of the fully coupled system have only appeared recently \cite{Ganopolski15aa}. 
 It would therefore be useful to determine the attractor properties associated with such  complex numerical systems forced by the astronomical forcing, and in particular estimate to what extent they may generate long transients and display sensitive dependence to noise.  The task involves both mathematical and technical challenges that will need to be addressed by steps of growing model complexity.

\section*{Appendix \label{table:params}}
The following sets of parameters are used in this study.
\begin{itemize}
\item{I98 model \cite{imbrie80}: $\tau =17$~ka, $b=0.6$, $L=0.249$, and $\epsilon =0.4$.}
\item{P98 model \cite{paillard98}: $X_i=0$, $X_g=1$, $X_G=1$, $\tau_i =10$~ka, $\tau_g =50$~ka, $\tau_G =50$~ka, $\tau_F =25$~ka, $i_0 =-0.75$, $i_1 =0$, $a=1$, $L=0.307$, and $\epsilon =0.4$.}
\item{CSW model \cite{Crucifix12aa}: $\alpha =30$, $\beta =0.75$, $\gamma =0.4$, $\tau =36$~ka, $L=0.810$, and $\epsilon =0.4$.}
\item{SM90 model \cite{saltzman90sm}: $p =1$, $q=2.5$, $r=0.9$, $s=1$, $u=0.6$, $v=0.2$, $\tau =10$~ka,  $L=0.546$, and $\epsilon =0.7$.}
\item{HA02 model \cite{hargreaves02}: $p=0.82$, $q=2.5$, $r=0.95$, $s=0.53$, $u=0.32$, $v=0.02$, $\tau =10$~ka, $L=0.726$, and $\epsilon =0.7$.}
\end{itemize}

\begin{acknowledgement}
MC is senior research associate with the Belgian National Fund of Scientific Research. This research is a contribution to the ITOP project, ERC-StG 239604 and to the Belgian Federal Policy Office project BR/121/A2/STOCHCLIM. 
\end{acknowledgement}
%
%
%
\bibliographystyle{spmpsci.bst}  
\bibliography{BibDesk.bib} 
\end{document}